\documentclass[
nofootinbib,aps,twocolumn,superscriptaddress,amsmath,amssymbm,showpacs
]{revtex4-1}

\usepackage{graphicx}
\usepackage{dcolumn}
\usepackage{bm}

\usepackage[percent]{overpic}

\usepackage{graphicx} 

\usepackage[usenames, dvipsnames]{color}

\usepackage{booktabs}

\makeatletter 
 
\newcommand{\ket}[1]{\left| #1 \right>} 
\newcommand{\bra}[1]{\left< #1 \right|} 
\let\baraccent=\= 
\renewcommand{\=}[1]{\stackrel{#1}{=}} 

  \renewcommand{\a}{\alpha}

  \renewcommand{\l}{\lambda}

 \newcommand{\s}{\sigma}
 \newcommand{\D}{\Delta}

\renewcommand{\k}{\kappa}

\newcommand{\E}{\mathcal{E}}

\renewcommand{\(}{\left(}
\renewcommand{\)}{\right)}

\newcommand{\norm}[1]{||#1||}

\usepackage{floatrow}
\usepackage[caption=false]{subfig}

\usepackage[normalem]{ulem}

\begin{document}

\author{Hao Jeng} \affiliation{Centre for Quantum Computation and
Communication Technology, Research School of Physics and Engineering,
The Australian National University, Canberra, Australian Capital
Territory 2601, Australia}

\author{Spyros Tserkis} \affiliation{Centre for Quantum Computation and
Communication Technology, School of Mathematics and Physics,
University of Queensland, St. Lucia, Queensland 4072, Australia}

\author{Jing Yan Haw} \affiliation{Centre for Quantum Computation and
Communication Technology, Research School of Physics and Engineering,
The Australian National University, Canberra, Australian Capital
Territory 2601, Australia}

\author{Helen M. Chrzanowski}
\affiliation{Centre for Quantum Computation and
Communication Technology, Research School of Physics and Engineering,
The Australian National University, Canberra, Australian Capital
Territory 2601, Australia}
\affiliation{Institute of Physics,
Humboldt-Universit\"at zu Berlin, Newtonstr. 15, D-12489 Berlin,
Germany}

\author{Jiri Janousek}\affiliation{Centre for Quantum Computation and
Communication Technology, Research School of Physics and Engineering,
The Australian National University, Canberra, Australian Capital
Territory 2601, Australia}

\author{Timothy C. Ralph} \affiliation{Centre for Quantum Computation and
Communication Technology, School of Mathematics and Physics,
University of Queensland, St. Lucia, Queensland 4072, Australia}

\author{Ping Koy Lam} \affiliation{Centre for Quantum Computation and
Communication Technology, Research School of Physics and Engineering,
The Australian National University, Canberra, Australian Capital
Territory 2601, Australia}

\author{Syed M. Assad} \email{cqtsma@gmail.com} \affiliation{Centre for
Quantum Computation and Communication Technology, Research School of
Physics and Engineering, The Australian National University, Canberra,
Australian Capital Territory 2601, Australia}

\title{Entanglement properties of a measurement-based entanglement
  distillation experiment}

\date{\today}
\begin{abstract}
  Measures of entanglement can be employed for the analysis of
  numerous quantum information protocols. Due to computational
  convenience, logarithmic negativity is often the choice in the case
  of continuous variable systems. In this work, we analyse a
  continuous variable measurement-based entanglement distillation
  experiment using a collection of entanglement measures.  This
  includes: logarithmic negativity, entanglement of formation,
  distillable entanglement, relative entropy of entanglement, and
  squashed entanglement. By considering the distilled entanglement as
  a function of the success probability of the distillation protocol,
  we show that the logarithmic negativity surpasses the bound on
  deterministic entanglement distribution at a relatively large
  probability of success. This is in contrast to the other measures
  which would only be able to do so at much lower probabilities, hence
  demonstrating that logarithmic negativity alone is inadequate for
  assessing the performance of the distillation protocol. In addition
  to this result, we also observed an increase in the distillable
  entanglement by making use of upper and lower bounds to estimate
  this quantity. We thus demonstrate the utility of these theoretical
  tools in an experimental setting.
\end{abstract}
\maketitle

\section{Introduction}
On the one hand, quantum entanglement is a useful non-classical
resource. It can be used for the construction of quantum gates
\cite{menicucci}, or for the distribution of cryptographic keys in a
secure manner \cite{scarani}. On the other hand, implementations of
these tasks are usually limited in performance due to experimental
imperfections. Utilising a variety of methods such as photon
subtraction \cite{takahashi, kurochkin} and noiseless linear
amplification \cite{chrzanowski, ulanov}, entanglement distillation
protocols seek a potential resolution to this problem by concentrating
weakly entangled states into subsets that are more entangled.

Here we address the problem of quantifying entanglement distillation;
this will, in general, depend on the kind of system that one is
working with. In the case of discrete variables, the fidelity with
respect to some maximally entangled state \cite{pan} and non-locality
based on the Bell inequalities \cite{gisin} are both useful measures
for observing quantum entanglement. However, these methods are not
particularly suitable in the case of continuous variables. Maximally
entangled continuous variable states do not exist, and a theorem due
to Bell precludes the demonstration of non-locality using Gaussian
states and Gaussian measurements (the standard tools for continuous
variable experiments) \cite{bell}, unless one introduces additional
assumptions on one's system \cite{oliver, ralph_bell}. Thus far, the
analyses of continuous variable entanglement distillation have instead
centred around inseparability criteria \cite{simon, duan} and, most
notably, the logarithmic negativity \cite{vidal} as one can calculate
it quite straightforwardly.

In this work, we analyse a continuous variable measurement-based
entanglement distillation experiment \cite{chrzanowski} using a
collection of measures. We present two main results. First, we show
that the logarithmic negativity is distinct from the other measures;
it crosses the ``deterministic bound'' before the other measures do,
at a probability of success (of the distillation protocol) that is
orders of magnitude greater. The deterministic bound is the maximum
entanglement that can be deterministically distributed across a given
quantum channel (usually imperfect), assuming that one had an EPR
resource state with infinite squeezing. For instance, when the
entanglement of formation crosses this bound, we can take it to
indicate a form of error correction \cite{spyros2}, thus giving an
example of how the logarithmic negativity can fail to capture
important properties of distillation protocols. Our results can be
regarded as an experimental demonstration of such an example.

Our second result is the certification of an increase in the
distillable entanglement. Currently, there is no known way for
evaluating the distillable entanglement directly, which means that one
is only able to look at it through the use of upper and lower bounds
\cite{andersen, andersen2}. The upper bound puts a limit on how much
distillable entanglement we had prior to distillation, while the lower
bound guarantees at least how much we have after performing
distillation. By observing a sufficient increase in the lower bound,
we could certify that the distillable entanglement has indeed
increased. We remark that the minimisation of optical loss and the
choice of a sharp upper bound turned out to be important factors in
order to observe the increase in the distillable entanglement. In
particular, there are many possible choices for the upper bound, but
the relative entropy of entanglement was found to be the only one that
was sufficiently stringent for this task.

We have organised this paper as follows. In section \ref{sec:pre}, we
provide some background in Gaussian quantum optics, and establish the
notations and conventions that will be used throughout this paper. In
section \ref{sec:measures}, definitions of the various entanglement
measures are provided, discussing their basic properties with an
emphasis on operational interpretations. In section \ref{sec:expt},
the final section, we briefly describe the experiment setup and
present a discussion of the distillation results based on the
measures.

\section{Preliminaries}
\label{sec:pre}
Gaussian states can be characterised by the first and second moments
of the quadrature operators \cite{weedbrook} --- also known
respectively as the mean fields and the covariance matrix. Since
measures of entanglement depend only on the covariance matrix, we will
assume vanishing mean fields without the loss of generality. For
two-mode Gaussian states, the covariance matrix can be written in
block form (using the $xpxp$ notation \cite{ukai}):
\[
\s=
  \begin{bmatrix}
    M & C \\
    C^T & N
  \end{bmatrix},
\]
where $M$, $N$, and $C$ are real $2 \times 2$ matrices. In general,
the entries of the covariance matrix depend on the value of $\hbar$;
in this paper, we will normalise to the variance of the vacuum field,
which amounts to setting $\hbar = 2$. In order for the covariance
matrix to represent a physical state, it is also required to satisfy
the Heisenberg uncertainty principle.

The density matrix of closed quantum systems evolve under unitaries:
$\hat{\rho} \rightarrow \hat{U}\hat{\rho}\hat{U}^\dagger$. In general,
representations of these unitaries in the Fock basis require
infinite-dimensional matrices; if we restrict ourselves to Gaussian
operations, this can be simplified to the evolution of covariance
matrices: $\s \rightarrow S\s S^T$. For two-mode states, each $S$ is a
$4 \times 4$ square matrix and is symplectic with respect to the
following symplectic form:
\[
\Omega=
  \begin{bmatrix}
    0 & 1 & 0 & 0 \\
    -1 & 0 & 0 & 0 \\
    0 & 0 & 0 & 1 \\
    0 & 0  & -1 & 0 \\
  \end{bmatrix}.
\]
It satisfies the equation $S \Omega S^T = \Omega$. Every Gaussian
unitary $\hat{U}$ is associated with a symplectic matrix. We will find
the following unitary useful:
\begin{equation}
\hat{S}_2(r) = e^{r(\hat{a}\hat{b}-\hat{a}^\dagger\hat{b}^\dagger)/2},
\end{equation}
which is known as the two-mode-squeezing operator. As usual, $\hat{a}$
and $\hat{b}$ denote the annihilation operators of the two optical
modes. The two-mode-squeezing operator is parametrised by the
squeezing parameter $r$, with $r = 0$ corresponding to no squeezing
and $r \rightarrow \infty$ to the limit of infinite squeezing.

Any given covariance matrix $\s$ can be put into the following
standard form:
\begin{equation}
\s=
  \begin{bmatrix}
    m & 0 & c_1 & 0 \\
    0 & m & 0 & c_2 \\
    c_1 & 0 & n & 0 \\
    0 & c_2 & 0 & n
  \end{bmatrix},
  \label{eq:sdf}
\end{equation}
and this can be done using only local Gaussian unitaries \cite{duan},
which does not change the entanglement of the state.

One can always diagonalise the covariance matrix using only symplectic
matrices, and the corresponding eigenvalues are called symplectic
eigenvalues \cite{williamson}. Of particular importance are the
symplectic eigenvalues of the partially transposed state. For two-mode
Gaussian states, the partial transpose flips the sign of the phase
quadrature, equivalent to flipping the sign of the $c_2$ entry in the
standard form of the covariance matrix (Eq.~\ref{eq:sdf}). The
symplectic eigenvalues $\tilde{\nu}_{\pm}$ of the partially transposed
state are given by
\begin{equation}
\tilde{\nu}_{\pm}^2 = \frac{\tilde{\D} \pm \sqrt{\tilde{\D}^2-4 \det
    \s}}{2},
\label{eq:symp}
\end{equation}
where $\tilde{\D} = \det M + \det N - 2 \det C$.

For convenience, covariance matrices of the form
\begin{equation}
  \s=
  \begin{bmatrix}
    m & 0 & c & 0 \\
    0 & m & 0 & -c \\
    c & 0 & m & 0 \\
    0 & -c & 0 & m 
  \end{bmatrix}
  \label{eq:sym}
\end{equation}
will be called symmetric, while those of the form
\begin{equation}
  \s=
  \begin{bmatrix}
    m & 0 & c & 0 \\
    0 & m & 0 & -c \\
    c & 0 & n & 0 \\
    0 & -c & 0 & n
  \end{bmatrix}
  \label{eq:qsym}
\end{equation}
will be called quadrature-symmetric. These states form strict subsets
of general two-mode Gaussian states (Eq.~\ref{eq:sdf}) by imposing
different kinds of symmetries.

\section{Entanglement measures}
\label{sec:measures}

We present some background on the theory of entanglement measures,
including definitions, properties, and formulae that will be
useful. We note that the problem of computing entanglement measures is
NP-complete for many measures \cite{huang}, with analytical
expressions available only in restricted cases that will generally
possess some degree of symmetry. In the special case of the
entanglement of formation, such a restriction will give rise to a
simple operational interpretation in terms of quantum squeezing, and
thus a connection to the logarithmic negativity which we briefly
discuss.

\subsection{Entanglement entropy}
The entanglement entropy $\E_V$ uniquely determines entanglement on
the set of pure states. Given the von Neumann entropy $S$, with
\begin{equation}
  S(\hat{\rho})= -tr(\hat{\rho}\log\hat{\rho}),
\label{eq:vN}
\end{equation}
the entanglement entropy $\E_V$ is defined to be the von Neumann
entropy of the reduced states \cite{bennett2}:
\begin{equation}
  \E_V(\ket{\psi}) = S(\hat{\rho}_A) = S(\hat{\rho}_B).
\label{eq:ent entropy}
\end{equation}
The subscripts $A$ and $B$ denote the two subsystems of $\ket{\psi}$:
$\hat{\rho}_A = tr_B(\ket{\psi}\bra{\psi})$ and
$\hat{\rho}_B = tr_A(\ket{\psi}\bra{\psi})$. For Gaussian states, the
von Neumann entropy depends only on the symplectic eigenvalues of the
covariance matrix. Throughout this paper, logarithms are taken with
respect to base $e$.

\subsection{Entanglement cost and distillable entanglement}
\label{subsec:eced}
\begin{figure}[t]
  \includegraphics[width = \textwidth]{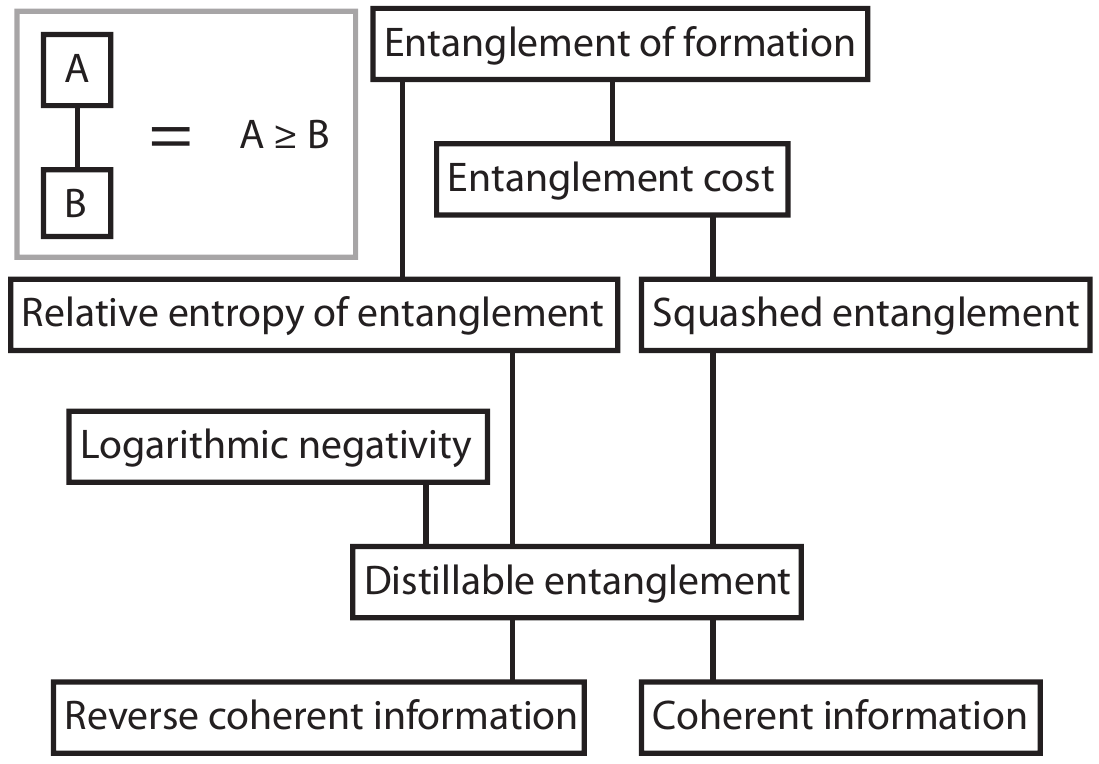}
  \put (-178, 143.5){\makebox[0.7\textwidth]{\cite{hayden,araki}}} 
  \put (-233, 114){\makebox[0.7\textwidth]{\cite{vedral}}} 
  \put (-163, 114){\makebox[0.7\textwidth]{\cite{christandl}}} 
  \put (-208, 68){\makebox[0.7\textwidth]{\cite{vedral}}} 
  \put (-163, 68){\makebox[0.7\textwidth]{\cite{christandl}}} 
  \put (-246, 54){\makebox[0.7\textwidth]{\cite{vidal}}} 
  \put (-195, 24){\makebox[0.7\textwidth]{\cite{devetak}}} 
  \caption{An ordering of the measures. If a measure $A$ sits above
    $B$ with a line connecting them, then
    $A(\hat{\rho}) \ge B(\hat{\rho})$ for all states $\hat{\rho}$. If
    one considers only pure states, then all measures shown in the
    diagram (except the logarithmic negativity) reduces to the
    entanglement entropy. When restricting to pure two-mode Gaussian
    states, the logarithmic negativity and the entanglement entropy
    can also be regarded as equivalent in the sense of
    Eq.~\ref{eq:conservation}. Citations in the figure refer to proofs
    for each inequality.}
\label{fig:ordering}
\end{figure}
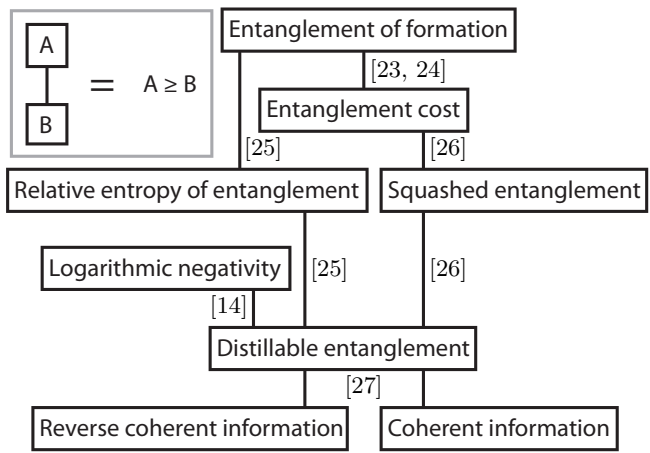

The extension of the entanglement entropy to mixed states is not
unique. Two possible choices are the entanglement cost and the
distillable entanglement; they are quite fundamental, since they
represent, respectively, the average pure state entanglement that is
needed for or that can be extracted from any given state (usually
mixed). The precise definitions are quite cumbersome to state and
somewhat unecessary for this paper, but they can be found in, for
instance, Horodecki \textit{et al.}  \cite{horodeckix4}. The
entanglement cost is an upper bound to the distillable entanglement
(Fig.~\ref{fig:ordering}), and they both reduce to the entanglement
entropy on the set of pure states. Neither measure is straightforward
to compute in general.

The entanglement cost is the asymptotic (regularized) version of the
entanglement of formation \cite{hayden}, which can be computed for
two-mode Gaussian states and is discussed in the next subsection. It
follows from this regularisation formula (and the sub-additivity of
the von Neumann entropy) that the entanglement of formation bounds the
entanglement cost from above \cite{hayden, araki}; in fact, recent
work shows that the two quantities coincide on a subset of states
\cite{wilde}, but the extent to which their equivalence holds is
presently unknown.

The distillable entanglement, on the other hand, does not admit any
closed expressions that we can evaluate straightforwardly. It can be
upper bounded by a number of quantities (see Fig.~\ref{fig:ordering}):
the entanglement of formation \cite{bennett}, the relative entropy of
entanglement \cite{vedral}, the squashed entanglement
\cite{christandl}, and the logarithmic negativity \cite{vidal}; it
also admits lower bounds due to the coherent and reverse coherent
information \cite{devetak}. We shall find such bounds useful for
estimating the distillable entanglement.

\subsection{Entanglement of formation}
\label{subsec:eof}
\floatsetup[figure]{style=plain,subcapbesideposition=top}
\begin{figure*}[t]
  \includegraphics[width = \textwidth]{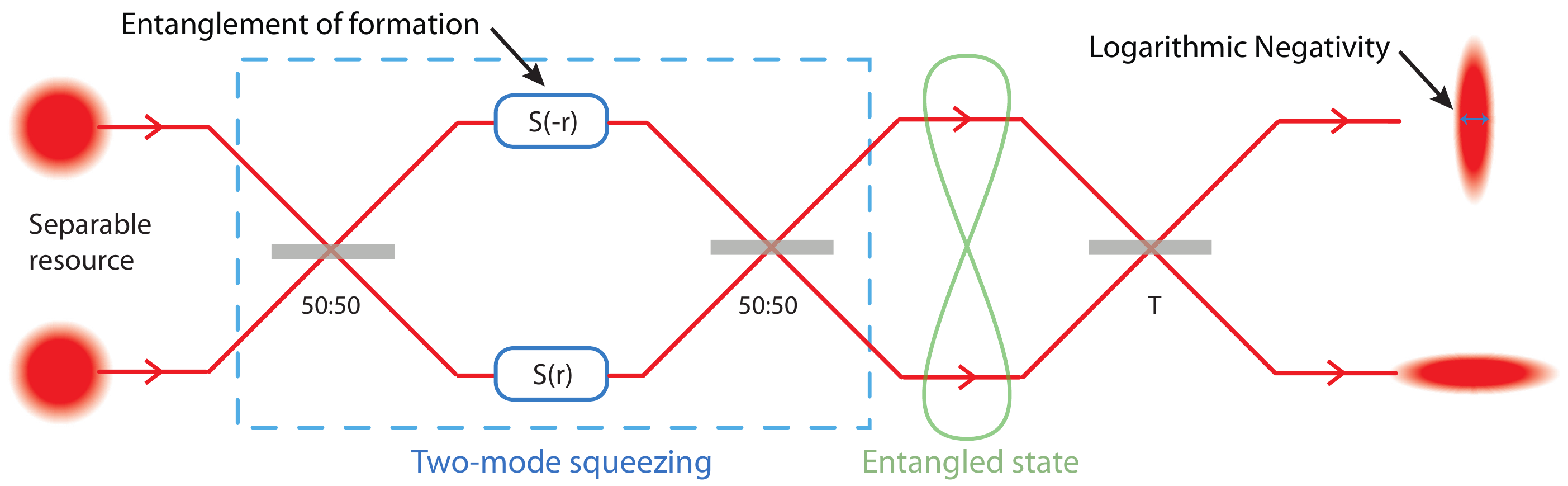}
  \caption{A comparison of the logarithmic negativity (equivalently,
    the PPT criterion) and the entanglement of formation. The
    operational interpretations of the two measures can be illustrated
    in terms of quantum squeezing, and the picture above holds good
    for two-mode Gaussian states that are symmetric in quadratures
    (those represented by Eq.~\ref{eq:qsym}). The minimum two-mode
    squeezing required to produce the entangled state corresponds to
    the entanglement of formation, while the maximum local squeezing
    that can be extracted from the entangled state can be identified
    with the symplectic eigenvalue $\tilde{\nu}_-$, and hence with the
    logarithmic negativity as well. The decomposition of two-mode
    squeezing into single-mode squeezers and passive operations has
    been shown explicitly; in addition, the beam-splitting ratio $T$
    that achieves maximal local squeezing will, in general, be
    state-dependent. The separable resource need not be symmetric
    (i.e.\ of the form in Eq.~\ref{eq:sym}) if the entangled state is
    not, and can also support non-vanishing correlations across the
    two modes as long as it remains separable.}
\label{fig:eof}
\end{figure*}
The entanglement of formation measures the minimum cost for producing
a state starting from pure entanglement resources \cite{bennett}:
\begin{equation}
  \E_F(\hat{\rho}) = \inf\left\{\sum_j p_j\E_V\(\ket{\psi_j}\) ~|~
    \hat{\rho} = \sum_j p_j \ket{\psi_j}\bra{\psi_j}\right\},
\end{equation}
where $\E_V$ is the entanglement entropy. The infimum runs over all
physical decompositions, including those that involve non-Gaussian
states; however, the minimum is attained by Gaussian states if
$\hat{\rho}$ is a two-mode Gaussian state \cite{akbari, ivan}. This
result also implies the equivalence between the entanglement of
formation and the \textit{Gaussian} entanglement of formation for
two-mode Gaussian states.

Unlike logarithmic negativity, the optimisation required by the
entanglement of formation makes computation difficult
\cite{giedke,marian}. A simple operational interpretation of the
entanglement of formation manifests when one restricts attention to
quadrature-symmetric states \cite{spyros}, allowing one to interpret
it as the squeezing operations required to produce the
state. Concretely, if $\sigma$ is a two-mode Gaussian state taking the
form of Eq.~\ref{eq:qsym}, then the entanglement of formation of
$\sigma$ is given by the following analytic expression:
\begin{equation}
  \label{eq:eof}
  \E_F(\sigma) = \cosh^2r_0 \log(\cosh^2r_0)-\sinh^2r_0\log(\sinh^2r_0),
\end{equation}
where
\begin{equation}
  \label{eq:r0}
  r_0 = \frac{1}{2}\log{\sqrt{\frac{\k-\sqrt{\k^2-\l_+\l_-}}{\l_-}}},
\end{equation}
with $\k = 2(\det \s + 1)-(m-n)^2$ and
$\l_\pm = \det M + \det N -2\det C + 2((mn+c^2)\pm 2c(m+n))$. The
meaning of $r_0$ is depicted in Fig.~\ref{fig:eof}--- it can be
identified as the minimum amount of two-mode squeezing that is needed
to produce the state $\sigma$, corresponding to an optimal choice of
the separable resource. For general two-mode Gaussian states, the
expression (Eq.~\ref{eq:eof} and \ref{eq:r0}) is a lower bound on the
entanglement of formation, and lies relatively close to the exact
value \cite{spyros}.

The symmetry requirements of Eq.~\ref{eq:eof} and \ref{eq:r0} is,
fortunately, not too stringent; for instance, the standard protocols
of entanglement swapping \cite{vloock} and entanglement-based quantum
key distribution \cite{cerf} work with entangled resources of this
type. However, experimental implementations of these protocols will
necessarily be imperfect, which means that quantum states produced in
the lab are never perfectly symmetrical.  In such cases, numerical
methods for calculating the entanglement of formation of arbitrary
two-mode Gaussian states can be quite useful \cite{wolf,adesso,
  marian}.

\subsection{Logarithmic negativity}
\label{subsec:logneg}
For arbitrary density matrices $\hat{\rho}$, the logarithmic
negativity $\E_N$ is defined to be \cite{vidal}:
\begin{equation}
\E_N(\hat{\rho}) = \log\norm{\hat{\rho}^{PT}}.
\label{eq:logneg}
\end{equation}
The symbol $\norm{\cdot}$ denotes the trace norm, the notation PT is
shorthand for the partial transpose. For two-mode Gaussian states
$\s$, the logarithmic negativity is a simple function of the
symplectic eigenvalue of the partially transposed state
$\tilde{\nu}_-$:
\begin{equation}
\E_N(\s) = 
\begin{cases}
0, \text{~if $\s$ is separable.}\\
-\log \tilde{\nu}_- , \text{~otherwise.}
\end{cases}
\end{equation}
It is thus a good indicator of inseparability by virtue of the PPT
(positive partial transpose) criterion \cite{peres,horodecki}, which
states that a Gaussian state is separable if and only if
$ \tilde{\nu}_- \ge 1$ \cite{simon}. The logarithmic negativity
coincides with the PPT-entanglement cost \cite{audenaert_ppt}.

While the symplectic diagonalisation of two-mode Gaussian states will
lead to uncorrelated thermal states, the symplectic diagonalisation of
its partial transpose will lead to squeezing. It is easy to see that
the maximum amount of local squeezing one can obtain from a two-mode
Gaussian state is given by the symplectic eigenvalue of its partial
transpose $\tilde{\nu}_-$, and that this can be achieved by
interfering the two modes on a beam-splitter
(Fig.~\ref{fig:eof}). Although it does not hold in the most general
case of Eq.~\ref{eq:sdf}, it does hold up to states with the
symmetries of Eq.~\ref{eq:qsym}. As a consequence of this operational
interpretation for $\tilde{\nu}_-$, it is related to the entanglement
of formation through the following inequality (ref.~\cite{adesso},
Eq. 43):
\begin{equation}
\label{eq:conservation}
\tilde{\nu}_- \ge e^{-2r_0},
\end{equation}
which essentially expresses the conservation of squeezing. The
equality is attained by symmetric states, but the two measures are in
general not equivalent. It has also been conjectured that $e^{-2r_0}$
is bounded from below by some non-trivial function of $\tilde{\nu}_-$
\cite{adesso}, and the gap between the upper and lower bounds would
imply that the two measures do not impose the same ordering on quantum
states. The disagreement of measures on the ordering of states holds
quite generally for the other entanglement measures as well
\cite{virmani}.

\subsection{EPR steering}
\label{subsec:steering}
By performing measurements of different observables on one party of an
entangled state, it is possible to steer the other party into
different types of quantum states
\cite{epr,schroedinger1,schroedinger2}. In continuous variable quantum
optics, EPR steering can occur when any of the following inequalities
on the conditional variances is violated \cite{reid}:
\begin{align*}
  &\E_{A > B} = V_{x_B|x_A} V_{p_B|p_A} \ge 1,\\
  &\E_{B > A} = V_{x_A|x_B} V_{p_A|p_B} \ge 1.
\end{align*}
The symbol $V_{x_A|x_B}$ denotes the conditional variance of the
quadrature $\hat{x}_A$ given $\hat{x}_B$, with the other quantities
interpreted in a similar fashion. The subscripts $A$ and $B$ denote
two parties sharing a bipartite entangled state. Two inequalities
instead of one is necessary for describing steering, because it is a
directional quantity. If we assume, without the loss of generality,
that entanglement is generated by the party A and distributed to the
party B, then we call $\E_{A>B}$ forward steering and $\E_{B>A}$
reverse steering. The party that performs the measurement is the party
that performs steering; that would be $A$ in the case of $\E_{A > B}$,
and $B$ in the case of $\E_{B > A}$. It is possible for a quantum
state to be steerable in one direction, but not in the other. In this
case, only one of the inequalities above is violated.

EPR steering is not a measure of entanglement, since it does not
characterise the separability of quantum states \cite{bowen, wiseman,
  vedral2}. It is a sufficient condition for inseparability, but not a
necessary condition --- there exists quantum states that are entangled
but not steerable in either direction. We note that EPR steering has
found applications in quantum key distribution --- in particular,
one-sided device-independent quantum key distribution \cite{branciard,
  walk} --- where the secure keyrate turned out to be a simple
function of EPR steering.

\subsection{Relative entropy of entanglement}
\label{subsec:ree}
The quantum relative entropy between any pair of density operators
$\hat{\rho}$ and $\hat{\s}$ is defined to be
\begin{equation}
  \label{eq:qrelent}
  S(\hat{\rho}||\hat{\s}) = tr\hat{\rho}(\log\hat{\rho}-\log\hat{\s}).
\end{equation}
The relative entropy of entanglement (REE) of $\hat{\rho}$ is then
defined by minimising the relative entropy over separable states
\cite{vedral2}:
\begin{equation}
  \label{eq:ree}
  \E_R(\hat{\rho}) = \inf_{\hat{\s}~\text{separable}}S(\hat{\rho}||\hat{\s}).
\end{equation}
By construction, it is zero if and only if $\hat{\rho}$ is
separable. The relative entropy of entanglement is an upper bound to
the distillable entanglement and a lower bound to the entanglement of
formation \cite{vedral}. One can further specialise the domain of
optimisation to Gaussian states, leading to the Gaussian relative
entropy of entanglement (GREE) \cite{scheel}:
\begin{equation}
  \label{eq:gree}
  \E_{GR}(\hat{\rho}) = \inf_{\substack{\hat{\s}~\text{separable} \\ \hat{\s}~\text{Gaussian}}}S(\hat{\rho}||\hat{\s}).
\end{equation}
The separable state which achieves the minimum of Eq.~\ref{eq:ree} is
called the closest separable state, and can be non-Gaussian even if
$\hat{\rho}$ is Gaussian \cite{wu}; hence, the relative entropy of
entanglement and its Gaussian approximation are not equivalent.

One does not have closed expressions for the relative entropy of
entanglement in general \cite{friedland}. Although there are numerical
methods based on semidefinite programming \cite{zinchenko}, this
technique is ill-suited in terms of computational time and memory
requirements for continuous variable systems; in this paper, we will
simply use the Gaussian relative entropy of entanglement as an
approximation. We show that the approximation is good in the regime
that we care about. Finally, we note that the relative entropy of
entanglement has been applied to the study of quantum repeaters; it is
an upper limit on the channel capacity, when one does not have access
to a quantum repeater \cite{pirandola}.

\subsection{Squashed entanglement}
Squashed entanglement is a measure based on the conditional mutual
information \cite{christandl}:
\begin{equation}
  \label{eq:sqsh}
  \E_{sq}(\hat{\rho}_{AB}) = \frac{1}{2}\inf_{\hat{\rho}_{ABE}} I(A:B|E),
\end{equation}
where one tries to minimise the conditional mutual information
$I(A:B|E) =
S(\hat{\rho}_{AE})+S(\hat{\rho}_{BE})-S(\hat{\rho}_E)-S(\hat{\rho}_{ABE})$
over all purifications $\hat{\rho}_{ABE}$ of the bipartite state
$\hat{\rho}_{AB}$. The subscripts $A$, $B$, and $E$ denote the
corresponding subsystems similar to that in Eq.~\ref{eq:ent
  entropy}. The optimisation is difficult to perform, but one can
exploit clever choices of the purification to obtain bounds on the
squashed entanglement \cite{goodenough}. Like the relative entropy of
entanglement, squashed entanglement is a bound on the capacities of
quantum communication channels; furthermore, it satisfies many axioms
of entanglement theory that other measures do not \cite{takeoka}.

\subsection{Coherent information}
The coherent \cite{schumacher} and reverse coherent information
\cite{garcia-patron} are defined as
\begin{align}
  \label{eq:cohinf}
  &I_C(\hat{\rho})= S(tr_A\hat{\rho}) - S(\hat{\rho}),\\
  &I_{RC}(\hat{\rho})= S(tr_B\hat{\rho}) - S(\hat{\rho}),
\end{align}
and, as usual, the $A$ and $B$ subscripts denote subsystems of the
bipartite state $\hat{\rho}$. These two measures are not entanglement
measures in the axiomatic sense \cite{vedral2}; however, they satisfy
the hashing inequality \cite{devetak, pirandola}:
\begin{equation}
  \label{eq:hashing}
  \max(I_C(\hat{\rho}),I_{RC}(\hat{\rho})) \le \E_D(\hat{\rho}),
\end{equation}
where $\E_D$ denotes the distillable entanglement. By virtue of the
hashing inequality, the coherent and reverse coherent information
provide sufficient conditions for inseparability, and plays the role
of a lower bound in characterising communication channels
\cite{pirandola}. This is in contrast to the relative entropy of
entanglement and squashed
entanglement, which would both correspond to upper bounds.\\

To conclude this section, we emphasise that we have made a choice to
work with one particular type of quantum correlation --- namely,
quantum entanglement. It is a suitable choice for the analysis of
entanglement distillation which we present in the next section. Other
interesting options include discord measures \cite{ollivier}, a
measure of squeezing \cite{idel}, and coherence measures \cite{tan};
however, we will not attempt to pursue these directions here.

\section{Experiment}
\label{sec:expt}
\subsection{Entanglement generation and processing}
\begin{figure*}[t]
  \includegraphics[width = \textwidth]{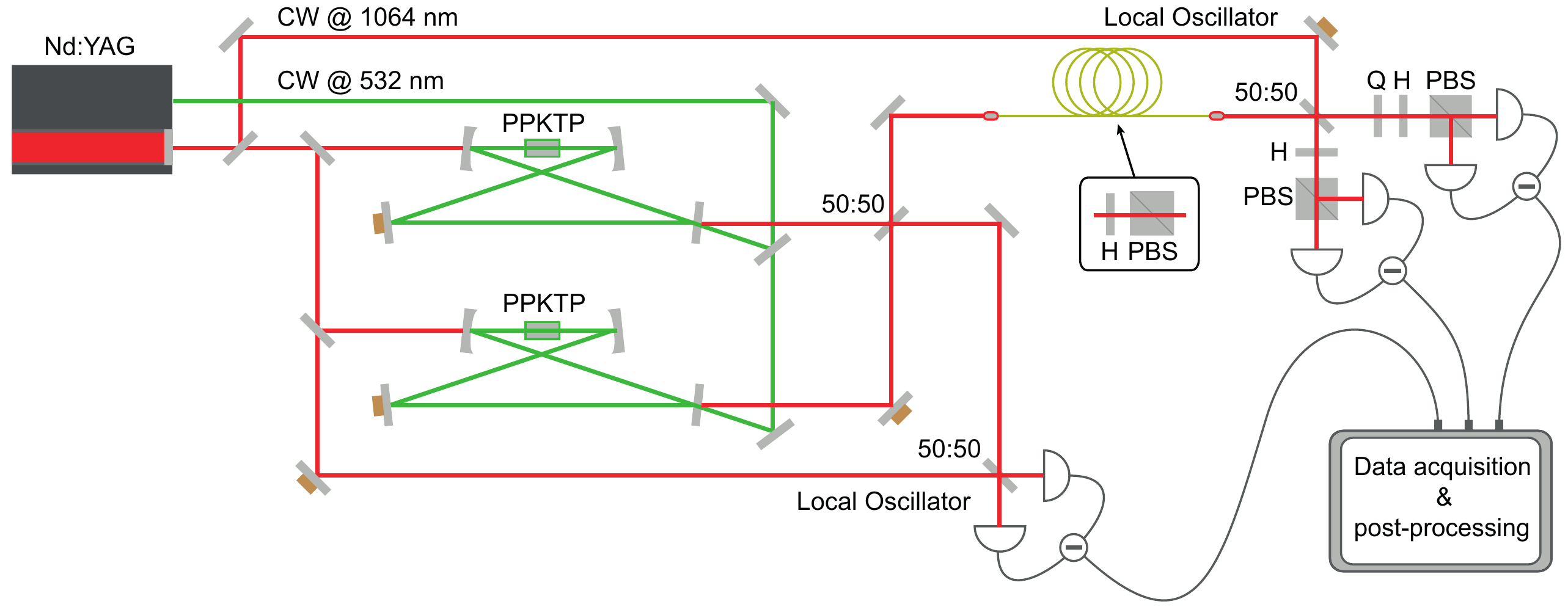}
  \caption{Schematic of the MBNLA experiment
    \cite{chrzanowski}. Squeezed states were generated from a pair of
    bow-tie cavities using the optical $\chi^{(2)}$ nonlinearity, and
    combined on a balanced beam-splitter to generate
    Einstein-Podolsky-Rosen entanglement. One of the beams from the
    EPR is sent through a communication channel that has optical loss,
    while the other is sent to a homodyne for verification; the
    transmission of the channel was characterised by optical
    heterodyne detection. Post-processing was applied on the data
    collected from the homodynes to emulate noiseless linear
    amplification, and thus the distillation of entanglement. H:
    half-wave plate, Q: quarter-wave plate, PBS: polarising
    beam-splitter, CW: continuous wave, PPKTP: periodically-poled
    potassium titanyl phosphate.}
\label{fig:mbnla}
\end{figure*}

We study measurement-based distillation of quantum entanglement using
recent advances in entanglement theory. The experiment setup
(Fig.~\ref{fig:mbnla}) consisted of a pair of bow-tie optical
parametric amplifiers driven at $532$ nm, with bright squeezed light
generated at $1064$ nm. The two beams were combined on a balanced
beam-splitter, {with a relative phase of $\pi/2$ to give
  Einstein-Podolsky-Rosen entanglement. One mode of this EPR pair was
  sent through a communication channel. Under the assumption that the
  quantum state is Gaussian, we can describe the state using just the
  first and second moments; five hundred sets of two million optical
  quadrature measurements were collected from each homodyne detector,
  retaining the 3-4 MHz narrowband through digital high-pass and
  low-pass filtering. Measurement-based entanglement distillation was
  performed using an approach similar to \cite{chrzanowski}, by
  post-processing the homodyne measurement data.


\begin{figure}[ht]
\includegraphics[width = \textwidth]{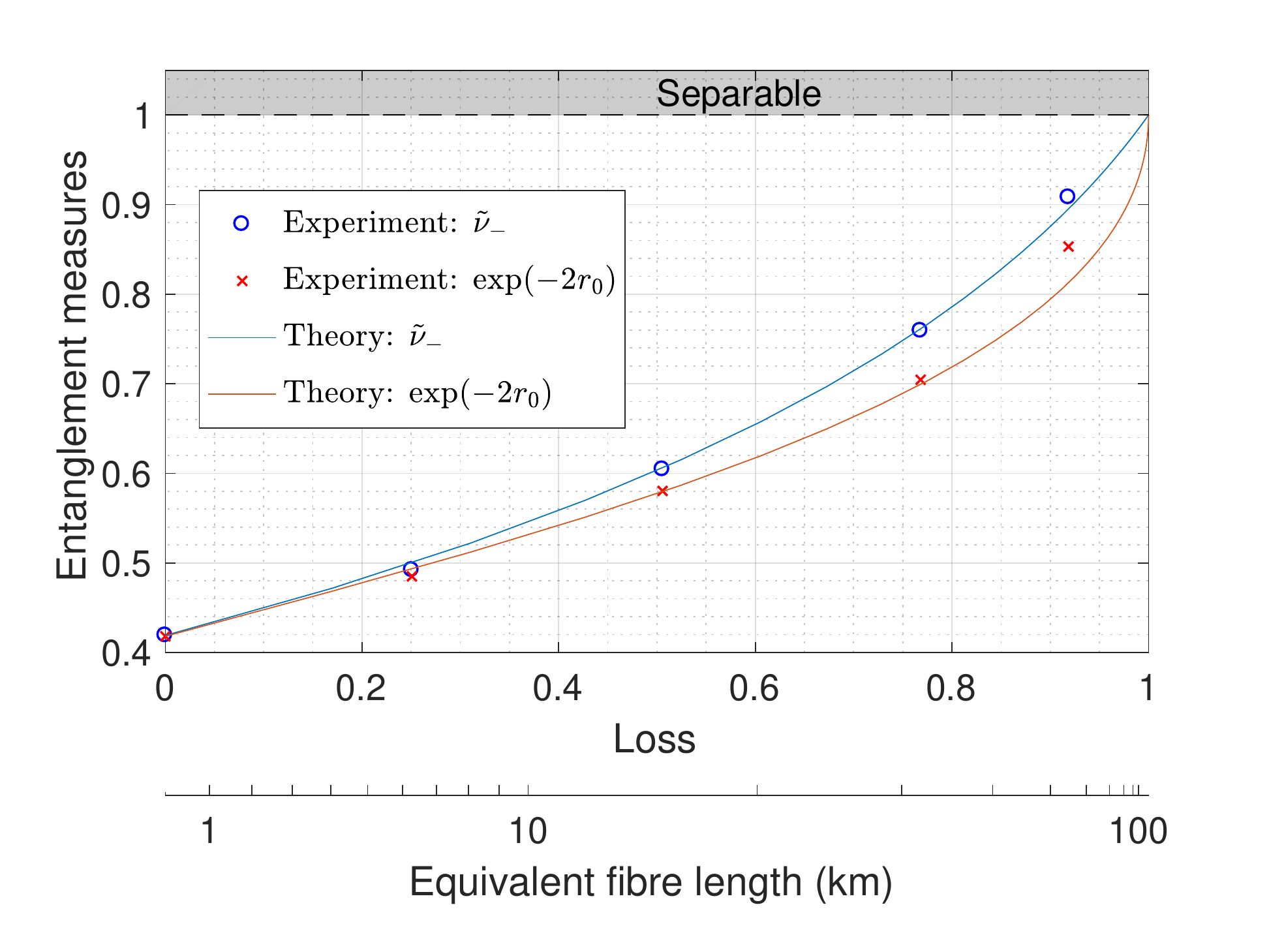}
\caption{Distribution of an entangled state through channels with
  various amounts of optical loss.  Circles are from experiments,
  solid lines are theory, and the shaded region above the dashed line
  correspond to separable states. We include a meter below the figure,
  which gives the loss-equivalent distance assuming telecom wavelength
  (i.e.\ 1550 nm), with the ticks distributed according to a
  logarithmic scale.}
\label{fig:loss}
\end{figure}

In practice, the communication channel through which we distribute the
entanglement will be lossy; here, we will assume that the loss is
entirely passive and model it as a beamsplitter with fixed
transmissivity. This was implemented using a polarising beam-splitter
preceeded by a half-wave plate. As we vary the loss, we can compare
the logarithmic negativity and the entanglement of formation using the
\textit{effective squeezing}, which varied according to
Fig.~\ref{fig:loss}. We use the term ``effective squeezing'' of the
logarithmic negativity and the entanglement of formation to refer to
the quantities $\tilde{\nu}_-$ and $\exp(-2r_0)$, respectively. These
quantities characterise the respective measures, and, at the same
time, can be interpreted as quantum squeezing (sections
\ref{subsec:eof} and \ref{subsec:logneg}.)  At unity transmissivity
(no loss), the state is symmetric but mixed, due to decoherence in the
entanglement generation process. Despite the mixture, the measures
remain equivalent due to the symmetry. In the presence of loss, the
states are asymmetric and the measures are no longer equal. This is
with the exception of maximal loss (zero transmissivity), where
nothing is transmitted and the state is separable. All measures
register effectively no squeezing in this case, which corresponds to
unity when expressed as a variance.

The effects of passive loss can be mitigated through the use of
noiseless linear amplification (NLA) \cite{ralph,fiurasek}. This
peculiar amplifier can be described by an unbounded operator
$g^{\hat{n}}$, where $g$ represents the amplitude gain. Acting the
noiseless linear amplifier on a coherent state will amplify the
complex amplitude without increasing the noise, while acting it on a
continuous variable EPR state with loss would lead to a state with
increased squeezing and reduced loss \cite{ralph}. Due to the
unbounded nature of the NLA operator, implementations are necessarily
approximate; in order to avoid violating the Heisenberg uncertainty
principle, they must also be non-deterministic
\cite{caves}. Experimental implementations of the NLA are subject to
additional limitations, such as restrictions on the size of the input
coherent amplitudes to small values \cite{pryde}.

In this paper, a virtual implementation of the noiseless linear
amplifier \cite{chrzanowski} has been chosen due to the ease of
implementation. An NLA followed by optical heterodyning is equivalent
to heterodyning followed by data processing; thus one is able to
implement the noiseless linear amplifier in the form of data
processing, provided that one performs a heterodyne
measurement. Concretely, one takes each outcome $\alpha$ of the
heterodyne and postselects it with an acceptance probability given by
\cite{fiurasek2}:
\begin{equation}
  \label{eq:filter}
  P(\a) = 
  \begin{cases}
    e^{(1-1/g^2)(|a|^2-|\a_c|^2)}, |\a| \le |\a_c|. \\
    1, |\a| > |\a_c|.
  \end{cases}
\end{equation}
where $g$ corresponds to the amplitude gain, and $\a_c$ is a constant
which specifies a cutoff. One then scales the successful events by
multiplication: $ \alpha \mapsto \alpha/g$. The postselection and
rescaling make up the data-processing stage which emulates the
noiseless linear amplifier. The closeness by which this
measurement-based implementation approximates the true NLA depends on
the cutoff --- a larger cutoff will improve the approximation at the
expense of a smaller probability of success \cite{ZJ}.

\subsection{Experiment analysis}
\begin{figure}[!ht]
  \subfloat{
    \includegraphics[width=\textwidth]{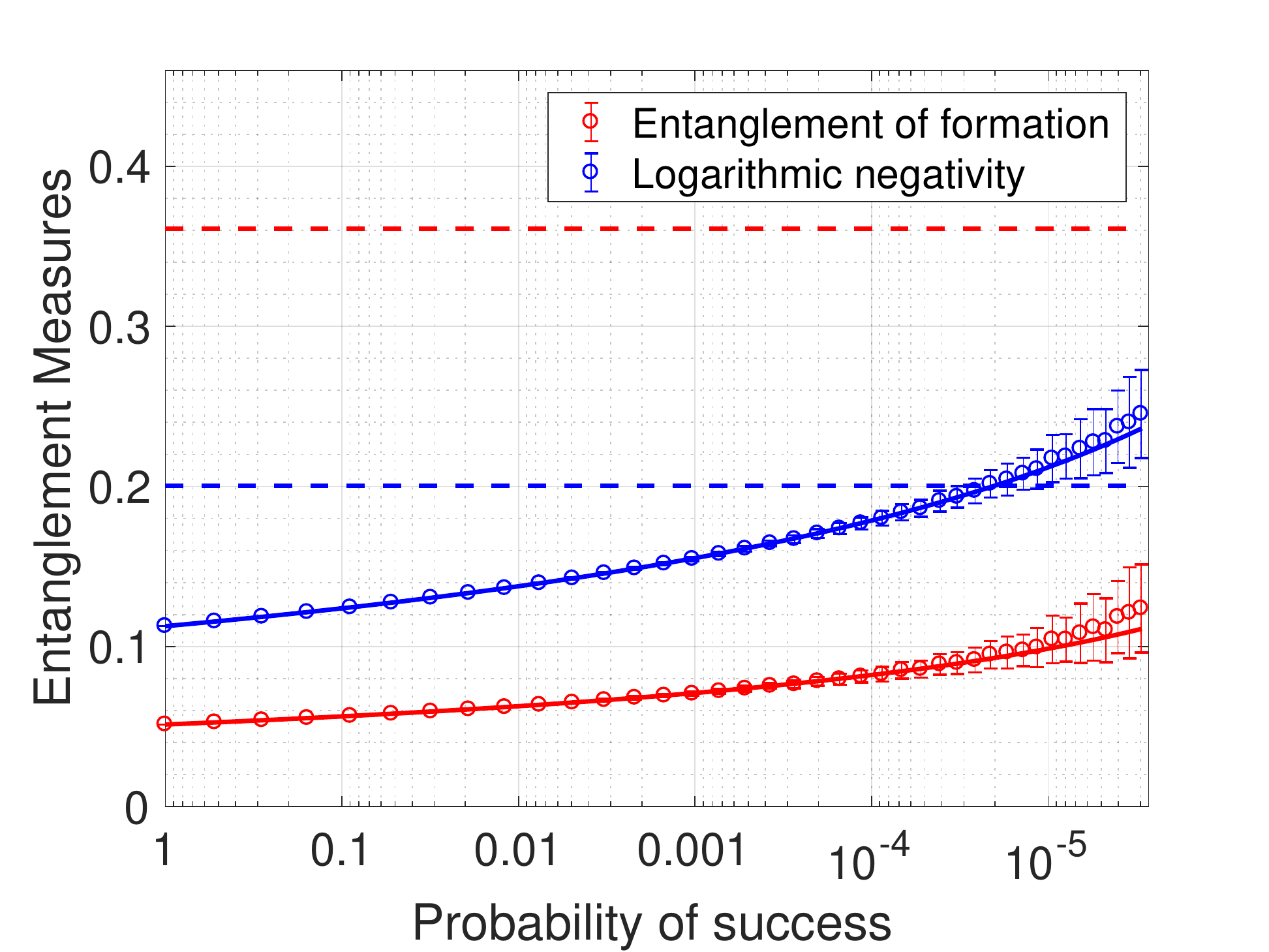}
    \put (-320, 165){\makebox[0.7\textwidth]{(a)}}
    \label{fig:em1}
  }\\
  \vspace{-3.5mm}
  \subfloat{
    \includegraphics[width=\textwidth]{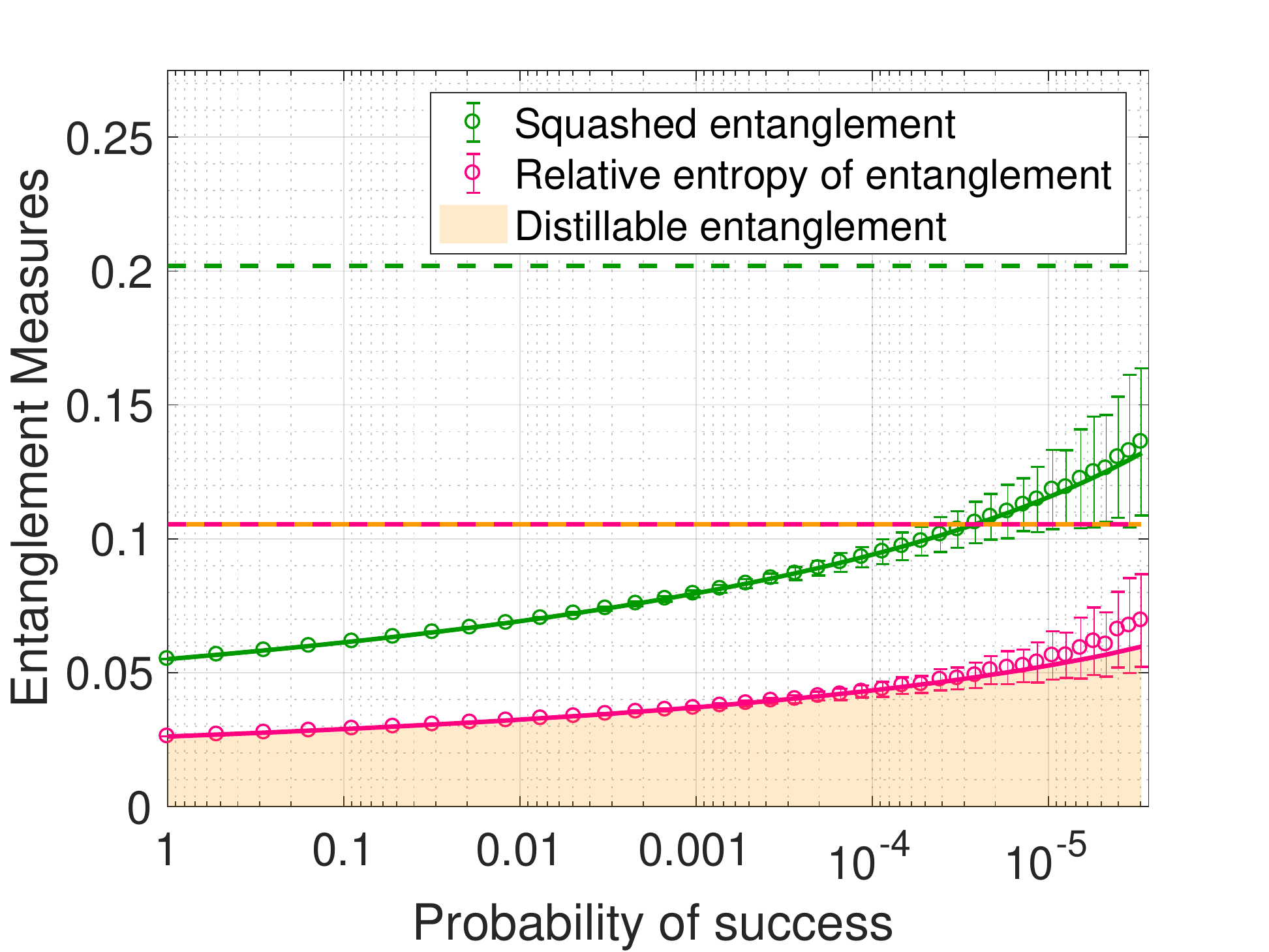}
    \put (-320, 167){\makebox[0.7\textwidth]{(b)}}
    \label{fig:em2}
  }
  \caption{Entanglement measures as a function of the probability of
    success. The optical loss was set to $90\%$. Circles correspond to
    experiment data, solid lines to theory calculations, dashed lines
    to the (measure-dependent) deterministic bound, and error bars
    represent 1.5 standard deviations over 100 repeated runs of
    postselection. The theory line can be calculated
    straightforwardly, assuming that the state is Gaussian (fitted to
    the measured covariance matrix) and that the postselection filter
    is ideal (i.e.\ infinite cutoff). The data points show positive
    bias relative to the theory model, which correspond to deviations
    from normality due to experimental imperfections and to the
    non-ideal filter implementation
    (Eq.~\ref{eq:filter}). \textbf{(a)} Both measures increase with
    decreasing probability of success, but only logarithmic negativity
    surpassed the deterministic bound. \textbf{(b)} All of the
    measures in this figure lie below their respective deterministic
    bounds. The distillable entanglement is bounded from above by the
    relative entropy of entanglement, and this is represented by the
    orange region; in the case of the deterministic bound, the two
    measures coincide (represented by the overlaying orange and
    rose-coloured dashed lines).}
\label{fig:em}
\end{figure}

\begin{figure}[ht]
  \includegraphics[width = \textwidth]{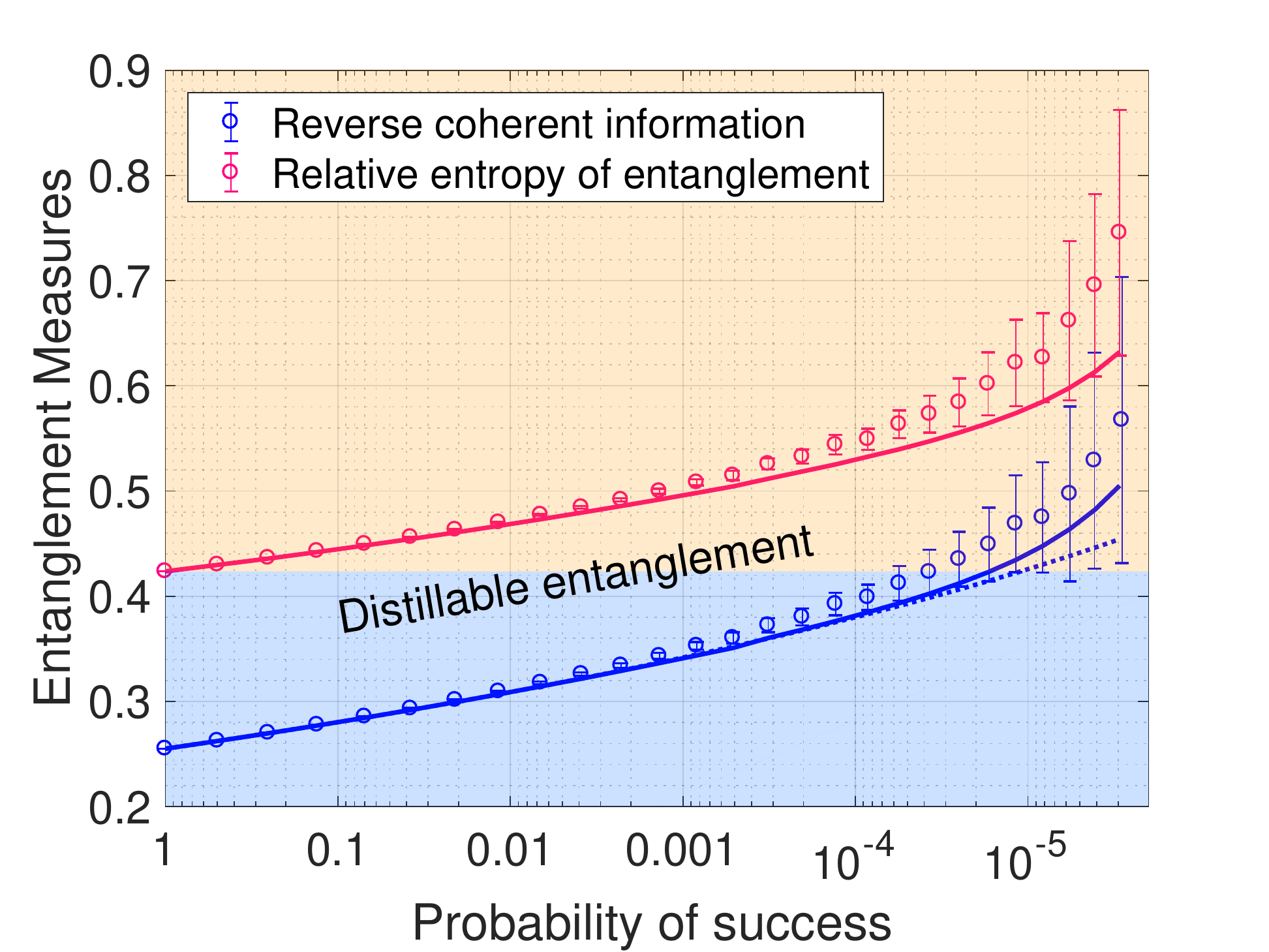}
  \caption{Demonstrating an increase in the distillable
    entanglement. The loss was set to $0\%$. The distillable
    entanglement is bounded from above and from below by the relative
    entropy of entanglement and the reverse coherent information,
    respectively. The boundary between the orange and blue regions is
    a horizontal line corresponding to the relative entropy of
    entanglement at unity probability of success, and values of the
    reverse coherent information lying in the orange region imply an
    increase in the distillable entanglement. The theory lines (given
    by the solid lines) in this figure assumes a finite cutoff, with
    the case of infinite cutoff also calculated for the reverse
    coherent information, showing that its values are smaller but
    remains in the orange region at small probabilities of success.}
\label{fig:em3}
\end{figure}

The results of the analysis is presented in Figs.~\ref{fig:em},
\ref{fig:em3}, and \ref{fig:notem}. We considered three
different settings of loss --- $90\%$ (Fig.~\ref{fig:em1},
\ref{fig:em2}, and \ref{fig:insep}), $50\%$ (Fig. \ref{fig:steer} and
\ref{fig:cohinf}), and $0\%$ (Fig. \ref{fig:em3}). For each loss, the
maximum gain $g$ for the postselection filter was set to $1.6$, $1.4$,
and $1.28$ respectively, with unity gain corresponding to no
postselection. We make the Gaussian assumption, and infer the
effective quantum state conditioned on successful postselection by
calculating the covariance matrix of the post-processed data. The
cutoff for the filter was chosen to be large enough to justify the
Gaussian assumption to at least $95\%$ using the Jarque-Bera test of
normality, which is based on skewness and kurtosis (the third and
fourth moments). The entanglement measures may finally be evaluated on
these effective states. We remark that different values of loss played
different roles --- a large amount of loss ($90\%$) draws a clear
distinction between logarithmic negativity and the other entanglement
measures; a moderate amount of loss ($50\%$) highlights the
directionality of EPR steering and of coherent information; and a
minimal amount of loss ($0\%$) allows us to certify an increase in the
distillable entanglement.

In Fig.~\ref{fig:em} and \ref{fig:notem}, all measures indicate
increasing entanglement with decreasing probability of success, as
they should. What is perhaps more interesting is a comparison with the
deterministic bound --- that is, the maximum entanglement that can be
transmitted through the channel in a deterministic fashion using an
EPR resource with infinite squeezing. The resulting state is known as
the Choi state of the channel \cite{choi}. We note that the
deterministic bound is measure-dependent, corresponding to the values
of each measure evaluated on the Choi state. In Fig.~\ref{fig:em1}, we
see that logarithmic negativity crosses its deterministic bound at a
relatively large probability of success, whereas the other measures
will also cross their respective bounds but at much lower
probabilities. The entanglement of formation was particularly far away
from the bound even at the small success probability of $10^{-6}$,
although it can in principle cross the bound at sufficiently low
probabilities of success \cite{spyros2}.  We attribute this
discrepancy to the operational meanings of the measures. The
entanglement of formation measures the squeezing operations needed to
produce an entangled state, while the logarithmic negativity is
related to local squeezing that can be extracted from the state. The
deterministic bound corresponds to a state for which a lot of
squeezing is needed to produce it, but not much can be extracted from
it; thus possessing a large entanglement of formation, but a lower
logarithmic negativity.

\begin{figure}[ht]
\includegraphics[width = \textwidth]{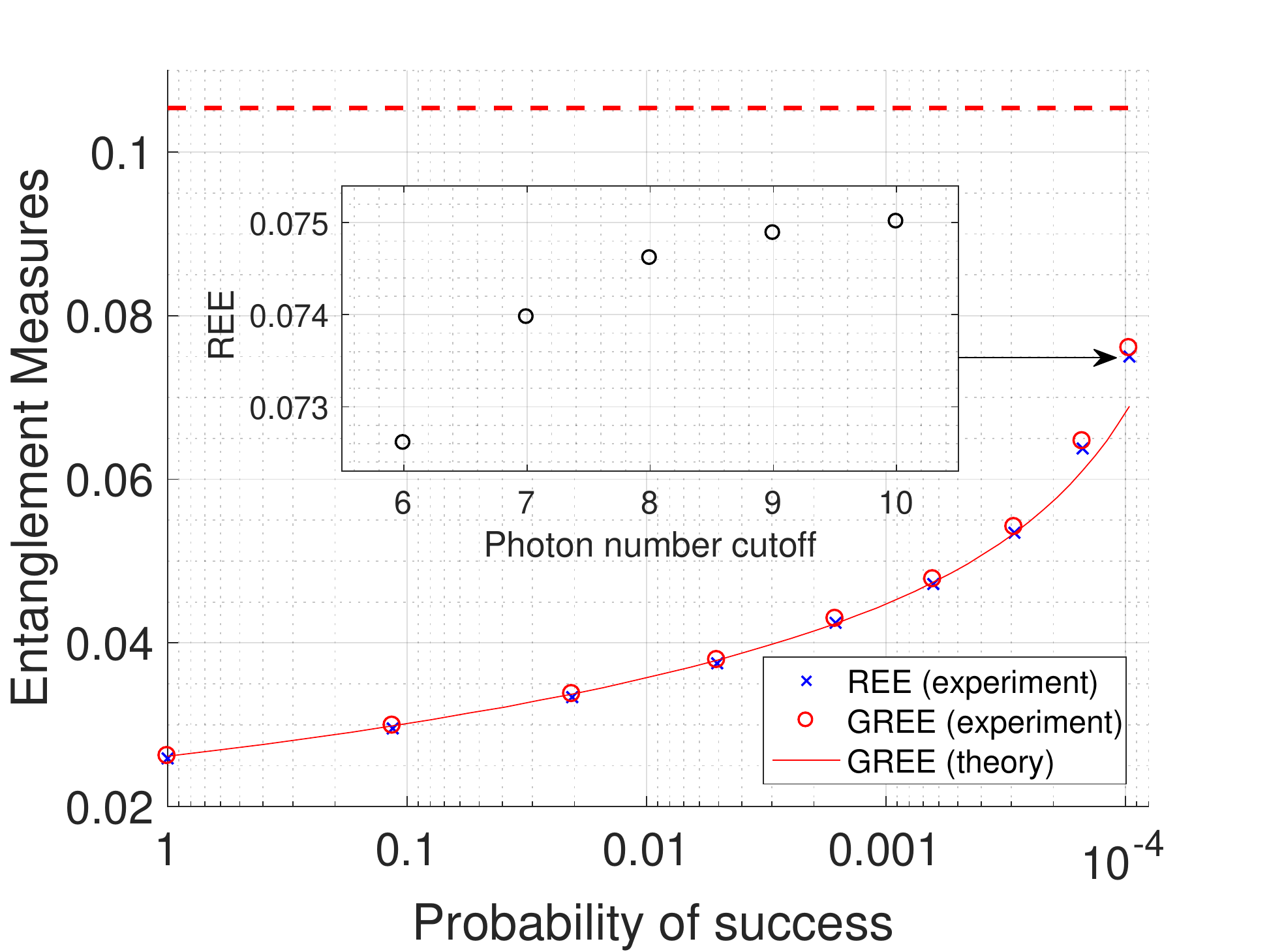}
\caption{A comparison of the relative entropy of entanglement and its
  Gaussian approximation. Inset shows the convergence of the numerical
  optimisation for the relative entropy of entanglement as a function
  of the photon number cutoff; the cutoff is necessary as we are
  approximating a continuous variable system using a
  finite-dimensional density matrix. The loss was set to $90\%$, and
  in such a case we find the two measures to be almost
  indistinguishable.}
\label{fig:ree}
\end{figure}

Results for the relative entropy of entanglement, for the squashed
entanglement, and for the distillable entanglement are shown in
Fig.~\ref{fig:em2}. We approximate the relative entropy of
entanglement using its Gaussian version (as explained in
Section~\ref{subsec:ree}), numerically performing the minimisation of
Eq.~\ref{eq:gree} over separable two-mode Gaussian states. This
approximation works relatively well when there is a large amount of
loss (Fig.~\ref{fig:ree}). Finally, the deterministic bound can be
calculated analytically \cite{pirandola}, and one finds that the
relative entropy of entanglement does not cross the deterministic
bound. We emphasise that the effects of noiseless linear amplification
--- the increase in squeezing and the reduction of loss \cite{ralph}
--- guarantees that any measure must cross the bound at sufficiently
small probabilities of success. However, the value of the probability
of success might be too small to be accessed in the experiment, which
is the case here.

Another important point to note for Fig.~\ref{fig:em2} is that the
deterministic bound for the relative entropy of entanglement coincides
with the bound for the distillable entanglement \cite{pirandola}. This
is a useful fact for showing that the distillable entanglement does
not cross the bound either. Although we cannot calculate the
distillable entanglement directly (for general states other than the
Choi state), we can bound it from above using the relative entropy of
entanglement, as illustrated by the orange shaded region in
Fig.~\ref{fig:em2}. This region lies below the deterministic bound,
thus demonstrating that the distillable entanglement does not cross
the bound.

We stress that the logarithmic negativity and the entanglement of
formation are unable to provide evidence of this, despite being upper
bounds of the distillable entanglement like the relative entropy of
entanglement is. Both measures cross the deterministic bound
\textit{that is given by the distillable entanglement}, which one can
read off Fig.~\ref{fig:em2} to be approximately 0.1 in value; thus
these measures are unable to rule out the possibility that the
distillable entanglement \textit{could have} crossed the deterministic
bound. As we have discussed in the previous paragraph, this cannot be
true because the distillable entanglement is always lesser than the
relative entropy of entanglement, which is in turn lesser than the
deterministic bound of the distillable entanglement. The conclusion
that the distillable entanglement did not cross the deterministic
bound can only be drawn using the relative entropy of entanglement as
the upper bound.

Fig.~\ref{fig:em2} also shows results for squashed
entanglement. Squashed entanglement is one of the measures for which
there exist no convenient methods for calculating it; at best, we have
a handful of bounds.  For the case of two-mode Gaussian states, one of
the best known bounds is given in \cite{goodenough}; it can be
evaluated for arbitrary phase-insensitive Gaussian channels, and hence
for states of the form Eq.~\ref{eq:qsym}, but not for those in the
general form of Eq.~\ref{eq:sdf}. We note that such a requirement can
be addressed by simply averaging the correlations of the amplitude and
the phase quadratures, which are sufficiently close for the two-mode
squeezed state generated in this experiment.  As shown in
Fig.~\ref{fig:em2}, squashed entanglement does not cross the bound;
however, one should keep in mind that the values for the squashed
entanglement are approximations in the form of an upper bound, and not
the actual value of the measure itself.

We pause briefly to discuss the significance of the collective results
in Fig.~\ref{fig:em}. The crossing of the deterministic bound by the
logarithmic negativity suggests that the distilled state is better
than the Choi state (i.e.\ an EPR state with \textit{infinite}
squeezing transmitted through the same communication channel). The
possibility of doing better than the Choi state is certainly not
forbidden by the laws of quantum mechanics, and can, for instance, be
achieved by using the noiseless linear amplifier with very large gains
\cite{ralph}. Nonetheless, we found that the logarithmic negativity
has apparently ``jumped the gun'' --- it suggests that this has
already been achieved in the present experiment when all the other
measures indicate otherwise. Thus, we have demonstrated a drawback for
using the logarithmic negativity as the figure of merit, which is
perhaps the price that one has to pay since it can be calculated so
trivially.
  
Returning to the analysis, we consider the distillable entanglement in
Fig.~\ref{fig:em3}.  It is similar to squashed entanglement in the
sense that neither can be evaluated using straightforward means, but
they differ because some of the bounds on the distillable entanglement
are rather stringent \cite{pirandola}. We employ these stringent upper
and lower bounds on the distillable entanglement to demonstrate an
increase in this quantity in the case of the lossless channel. We note
that such a task would have been trivial if we knew how to calculate
the measure --- the need for using bounds in the case of the
distillable entanglement is precisely because there is no method for
calculating it directly.

Similar to Ref.~\cite{andersen,andersen2}, we use the reverse coherent
information to bound the distillable entanglement from below; however,
we use the relative entropy of entanglement instead to bound it from
above (as opposed to using the logarithmic negativity in
Ref.~\cite{andersen, andersen2}). If, after performing the
entanglement distillation, the reverse coherent information ends up
greater than the relative entropy of entanglement that we had started
off with (indicated by the orange shading in Fig.~\ref{fig:em3}), one
may conclude that the distillable entanglement has increased. If the
reverse coherent information remains smaller, then no conclusion can
be drawn (corresponding to the blue shading). Fig.~\ref{fig:em3} shows
values of the reverse coherent information surpassing the bound given
by the relative entropy of entanglement at low probabilities of
success, and hence an increase in the distillable entanglement. We
remark that these results are fundamentally limited by state
preparation and measurements --- it cannot be improved simply by
adjusting the postselection settings, due to excess noise and to the
diminishing probability of success. In addition, the upper bound and
the channel transmissivity cannot be arbitrary. There are other
choices for the upper bound (Sec.~\ref{subsec:eced}), and one can also
consider other settings for the loss (anything between $0\%$ and
$100\%$); however, no increase in the distillable entanglement was
observed in any of these cases using the method presented above. In
this sense, the results in Fig.~\ref{fig:em3} is optimal.

\begin{figure}[!h]
  \subfloat{
    \includegraphics[width = \textwidth]{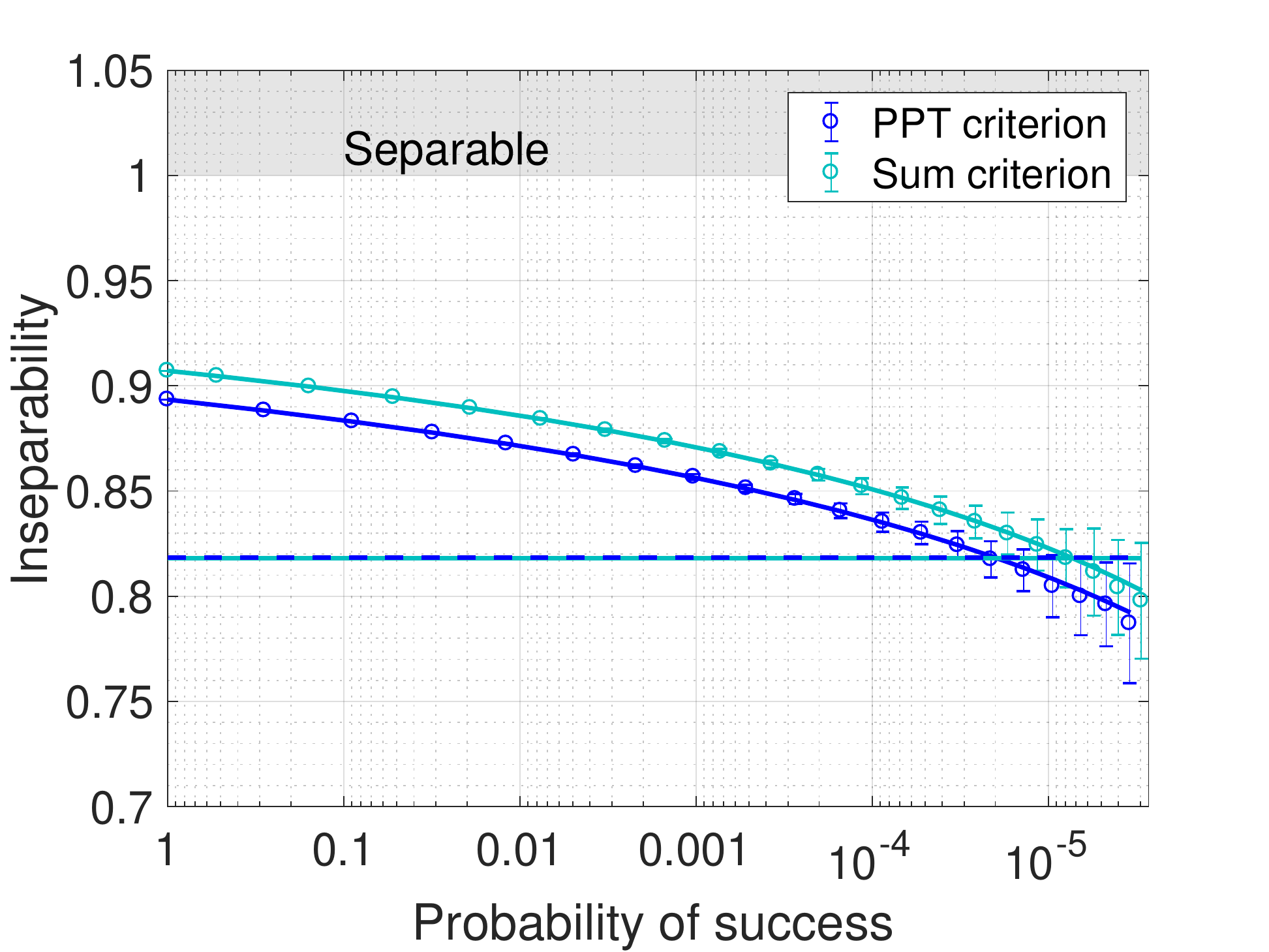}
    \put (-325, 160){\makebox[0.7\textwidth]{(a)}}
    \label{fig:insep}
  }\\
  \vspace{-3.5mm}
  \subfloat{
    \includegraphics[width = \textwidth]{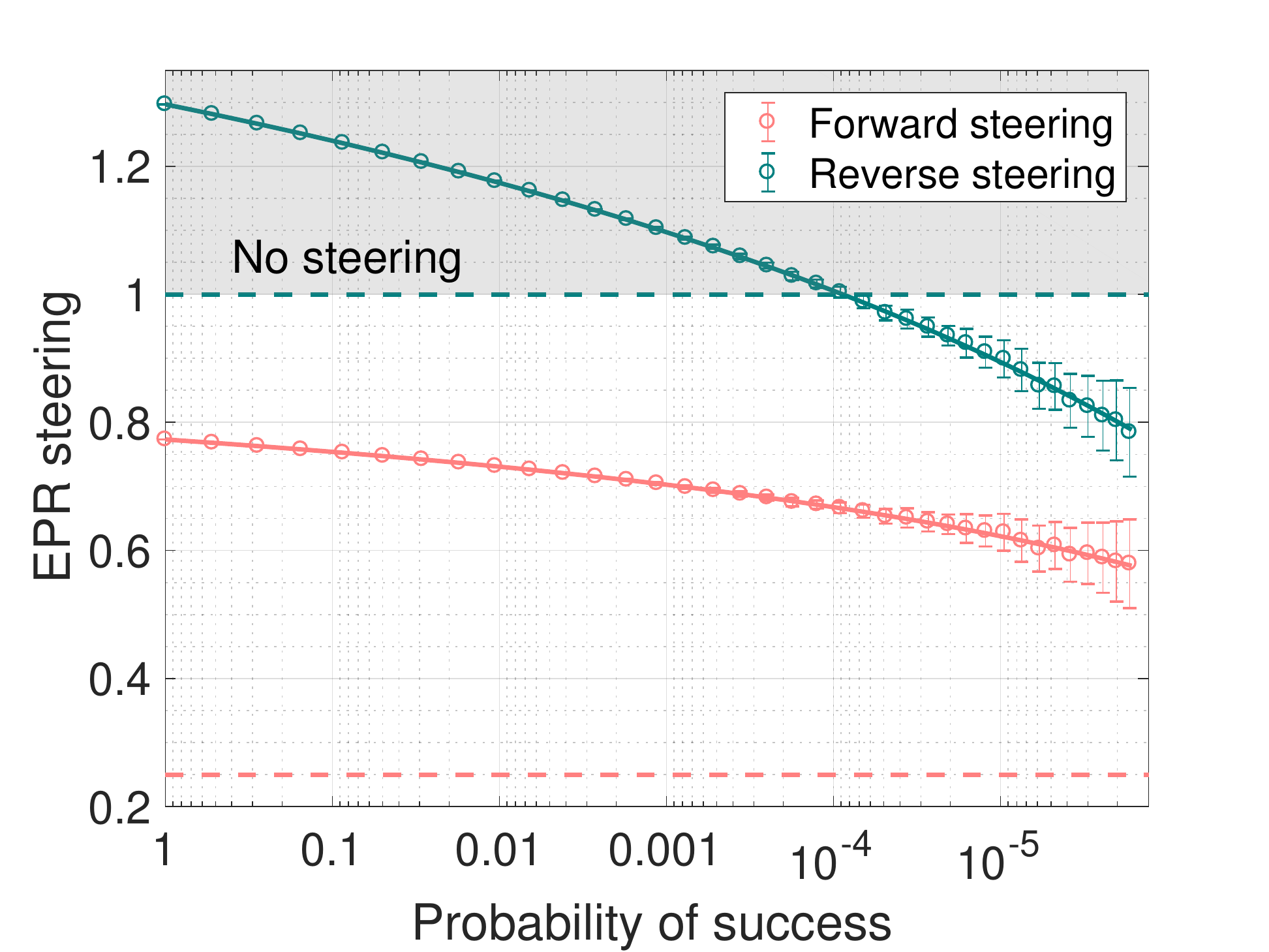}
    \put (-325, 165){\makebox[0.7\textwidth]{(b)}}
    \label{fig:steer}
  }\\
  \vspace{-3.5mm}
  \subfloat{
    \includegraphics[width = \textwidth]{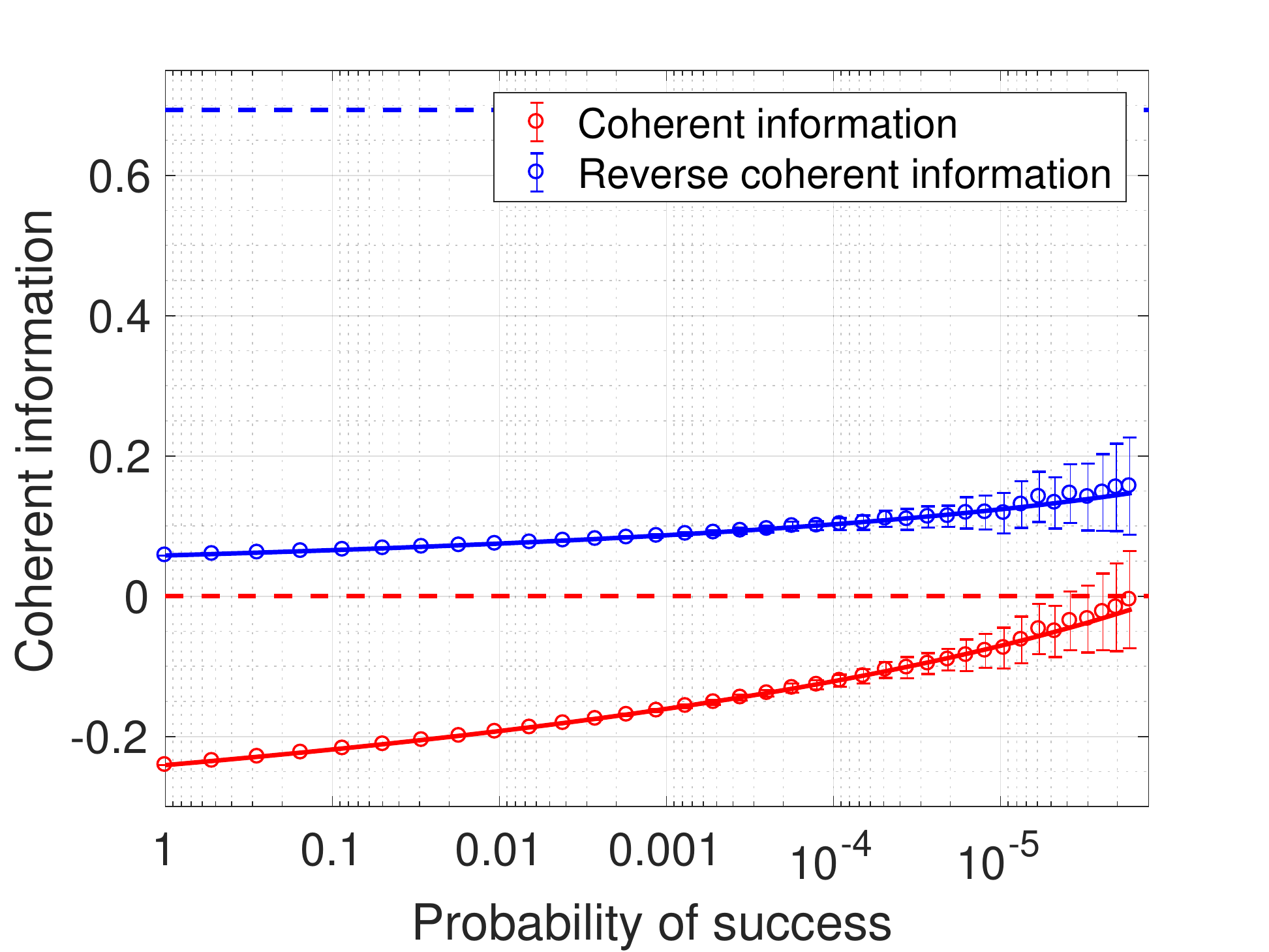}
    \put (-325, 165){\makebox[0.7\textwidth]{(c)}}
    \label{fig:cohinf}
  }
  \caption{Relatives of entanglement measures. The loss in \textbf{a}
    was set to $90\%$, while that in \textbf{b} and \textbf{c} is
    $50\%$. The data points, theory lines, deterministic bounds, and
    the error bars should be interpreted in the same way as those for
    Fig.~\ref{fig:em}. Noiseless linear amplification is useful for
    recovering forward steering and coherent information, but less so
    for the other direction which is more robust to loss.}
\label{fig:notem}
\end{figure}

Adding to the collection, we consider relatives of entanglement
measures. These are not proper entanglement measures in the axiomatic
sense. Fig.~\ref{fig:insep} illustrates the inseparability criteria
for two-mode Gaussian states; the sum criterion \cite{duan} displays
similar behavior to the PPT criterion, as both measures deal with the
extraction of squeezing from entangled states \cite{giovannetti}. The
sum criterion relies on an extraction protocol that is suboptimal
compared to the PPT criterion, hence its values are closer to the
separable boundary. Both are, of course, equally valid for certifying
inseparability.

In Fig.~\ref{fig:steer} we show the results for EPR steering, a
directional quantity for which the properties depend on the direction
of interest. Reverse steering is particularly succeptible to loss,
where an entangled state transmitted through $50 \%$ of loss would not
be steerable in the reverse direction. It is interesting to note that,
for this particular case, violation of the EPR steering criterion is
equivalent to surpassing the deterministic bound. This is not true in
general. By performing noiseless linear amplification, one is able to
recover reverse steering beyond the deterministic bound at reasonable
success probabilities, and one may compare this with forward steering
--- the deterministic bound is smaller to begin with, and is also much
harder to beat. Reverse steering is sensitive to loss because one is
trying to steer using the lossy mode --- most of the information about
the entangled state has already been lost to the environment. In the
case of direct steering, there is an advantage since one is using the
mode that does not suffer from the loss of the channel.

Finally, we consider the coherent information and the reverse coherent
information, which are related to entanglement through the hashing
inequality (Eq.~\ref{eq:hashing}). The reverse coherent information is
robust against loss for the same reason that direct steering is;
likewise, the coherent information is fragile the same way reverse
steering is succeptible to loss (Fig.~\ref{fig:cohinf}). The reverse
coherent information is always positive even when no noiseless linear
amplification was performed, while the coherent information was not
initially positive but could be recovered at some small probability of
success that is just out of reach in this experiment. Due to the
robustness of the reverse coherent information, the deterministic
bound is much harder to surpass.

\section{Conclusion}
By analysing a measurement-based entanglement distillation experiment
using a collection of measures, we showed that the logarithmic
negativity exhibits behavior quite distinct from the others. It would
make us believe that more entanglement has been distilled than what is
offered by the deterministic bound, in stark contrast to what the
other measures suggest. In addition to this result, we were also able
to certify an increase in the distillable entanglement (in the case of
the lossless channel), relying primarily on a judicious choice of the
upper bound in order to estimate this quantity accurately. The work we
have presented is useful for analysing entanglement distillation, but
can also be extended to more general situations; this includes
entanglement swapping, for instance, and the analysis of quantum
repeaters in general.

\begin{acknowledgments}
  This work is supported by the Australian Research Council (ARC) under the Centre of Excellence for Quantum Computation and Communication Technology (CE110001027, CE170100012, FL150100019).
\end{acknowledgments}

\end{document}